# A totally non-compensatory multi-criteria method for evaluating and improving level of satisfaction (LoS): proposal and application on Airport Terminal of Passengers

**Phelipe Medeiros da Rocha**
Universidade Federal Fluminense
Rua Passos da Pátria, 156
Niterói, RJ , Brazil, CEP 24.220-045
`pheliperocha@id.uff.br, phelipe_m_rocha@yahoo.com.br`

Helder Gomes Costa
Universidade Federal Fluminense
Rua Passos da Pátria, 156
Niterói, RJ , Brazil, CEP 24.220-045
`heldergc@id.uff.br` [Corresponding author]

## Abstract

To evaluate and assign a service according customer's level of satisfaction (LoS) is a relevant issue in operations management. This is a typical situation in which the evaluators, have passed by heterogeneous experiences along their life which implies they could consider different variables when evaluating a product. Despite it, the models for measuring Los usually consider a homogeneous set of criteria when facing LoS evaluation. This study applies a totally non-compensatory modeling that allows each customer to select the criteria, from a whole set of aspects, the customer wants to use for evaluating LoS. The proposal was tested in evaluating LoS regarding the services provided by Airport Terminal of Passengers (ATPs) in Brazil, with data collected in a survey involving 19,240 passengers, interviewed at 15 Brazilian international airports. The data collected was imputed into ELECTRE TRI ME algorithm to obtain the a credibility degree of sorting the instances. The values of credibility degree were them used to obtain groups of ATPs. Finally, the statistical modes of the evaluations in each group were analyzed and compared. The proposal allowed a full non-compensatory approach to obtain the credibility degree even when considering perceptions from several evaluators that could use different criteria. As a result, it was identified, for each cluster of ATP, the criteria sets to be improved and even those to be prioritized. The pioneer modeling proposed in this article for evaluating LoS plus its instancing in ATPs terminals represents an original advance in the establishment of a multi-criteria decision aid (MCDA) model to assess the quality of services and fills a relevant gap for a full non-compensatory approach able to classify the LoS in the airport context, considering perceptions of multiple evaluators even if they use different criteria in their evaluations.

# 1 Introduction

In the environment of industrial and operations management, one of the most important issues to consider is how to evaluate and assign the quality perceived by consumers. As stated by Francis et al. [2003], the Level of Satisfaction (LoS) is a term frequently used as a proxy to assess the quality of a service (QoS) or a product.

The LoS evaluation presents the following typical features:

- **[Feature 1]**. The customer takes into account a variety of factors during evaluation. So, as you can see in Table 1, the evaluation of LoS and multi-criteria decision aid (MCDA) have some features in common.
- **[Feature 2]**. In several LoS evaluations, compensation for performance is not admissible. For instance, a beautiful appeal of a meal should not compensate its very poor taste.
- **[Feature 3]**. An important aspect regarding the use of MCDA for approaching LOS evaluations is the presence of several evaluators. It is an issue, because MCDA techniques were not primarily developed to deal with multiple evaluators.
- **[Feature 4]**. The evaluators, who are typically consumers, have gone through a variety of experiences throughout their lives, which means that each evaluator should have its own set of variables or aspects when evaluating a product. Despite this, the models that are used to measure customer satisfaction typically consider a comparable set of factors in order to obtain the customers' perceptions regarding the performance of a product.

As a consequence of the **Feature 1**, MCDA-based approaches have been stated to classify the LoS based on customers' evaluations across viewpoints or criteria. These developments have taken two major paths, depending on whether we consider multi-attribute utility theory (MAUT) or outranking approaches.

Table 1: Multi-criteria decision and level of satisfaction evaluation: comparing the main features

|              | Multi-criteria Decision Problems | LoS evaluation           |
|--------------|----------------------------------|--------------------------|
| Variables    | More than one                    | More than one            |
| Evaluations  | Subjective                       | Subjective               |
| Problem type | Decision                         | Evaluation and Assessing |

Table 15 in Appendix A provides a non-exhaustive list of prior models that utilize MAUT methods to assess the LoS. The most commonly used MAUT approaches in this field are AHP [Saaty, 1980], TOPSIS [Hwang and Yoon, 1981], MUSA [Siskos and Grigoroudis, 2002], and DEMATEL [Gabus and Fontela, 1973].

While these works have made worthy advances, they are open to criticism due to the sensibility of MAUT models to compensatory factors. For instance, a high performance under a criterion should compensate or balance a low performance in another criterion. This means that they have no protection against the issue connected to **[Feature 2] of LoS**.

On the other hand, as one can see in Table 16 in Appendix A, there are modeling based on applying MCDA outranking methods that were constructed to evaluate LoS and addressed to avoid compensatory effects.

But these models were not designed for handling evaluations from more than one person (**Feature 3** of LoS) because they first use compensatory functions of preference to add up the evaluations and then use the outranking method. The problem arises from the common practice of using the arithmetic mean for preliminary processing evaluations, which introduces compensatory effects in the data processing.

This critique was addressed and handled by Costa and Duarte [2019] and da Rocha and Costa [2021] who employed the ELECTRE TRI ME method [Costa et al., 2020] specifically designed to exclude any compensatory effects in scenarios involving multiple evaluators. Even though there have been improvements, Costa and Duarte [2019] and da Rocha and Costa [2021] looked at situations to judge one case, which means they suggested ways to make each case better. Another gap not covered in the works of Costa and Duarte [2019] and da Rocha and Costa [2021] is related to **[Feature 4 of LoS**, as they did not consider different evaluators could have considered different sets of criteria.

We conclude that, in spite of the great contributions and advances provided by applying MCDA for assessing LoS, there is a gap for a model with the following properties:

- Uses a non-compensatory multiple criteria sorting technique.
- Does not apply compensatory compensatory functions do pre-process the data.

- Applied in a situation having several instances to be sorted.
- Regards heterogeneity of evaluators - evaluators do not need to use the same whole set of criteria.

### 1.1 Objective and highlights about the proposal

The main goal of this study is to contribute to filling this gap by creating and using a new model for evaluating LoS, which is based on pure outranking and multiple evaluators. In addition to addressing the gap described early in this Introduction section, we highlight the following features of our proposal:

1. It prevents any compensatory effects when combining evaluations from a sample of individuals with diverse life experiences and expectations. For instance, it approaches the following question: if one customer rates a service as "Amazing" and another user evaluates the same service as "Terrible", should the service be characterized as having a "Regular" or "Median" performance?

2. It enables many evaluators to assess a service using distinct criteria. This feature addresses the challenges associated with circumstances where the client, based on their own perception, does not feel adequately informed to assess a certain component of the service.

   For instance, let's consider a scenario where a customer in a shopping mall is requested to assess the quality of the parking facilities while not having utilized the parking space. Suppose the consumer declines the request to evaluate the parking services but is willing to assess other parts of the shopping mall's services as an example.

   What is the appropriate course of action when seeking a comprehensive assessment of the service? Should we discard all the evaluations from this customer? Should we exclude the parking conditions factor from the study, disregarding all the evaluations made by the other respondents? Should we populate the empty cell with the average value of the responses provided by the other participants in that specific category?

3. It addresses situations where the service fails to demonstrate consistent performance across all factors or criteria that the sample evaluates. For example, let's consider a scenario where a consumer visits a restaurant and rates the parking area service as "excellent", but has a negative perception of the staff's service quality, considering it to be "poor".

## 2 Multicriteria outranking methods applied for LoS evaluation

There are previous works that use a multi-criteria-based approach to classify the LoS services according to evaluators' or customers' perceptions regarding their performances under multiple aspects, viewpoints, or criteria. Some of them are based on MAUT methods, as for instance: Pamucar et al. [2021], Bezerra and Gomes [2020], and da Rocha et al. [2016]. Conversely, other studies have implemented outranking methods.

As we focus on developing a model capable of avoiding compensatory effects, in this section, we highlight previous works that have utilized outranking modeling for QoS evaluation.erefore, it follows non-exhaustive comments about such previous contributions, highlighting their main feature.

As far as we found out, Freitas and Costa [1998] was a pioneer in proposing the use of MCDA for the evaluation of QoS. This paper suggested using ELECTRE III (Roy [1978]), a program that was created to help make decisions when there are multiple points of view. It does this by using an outranking method to avoid compensatory effects that aren't desirable. Such a proposal, named by ELEQUAL, was applied to evaluate the services provided by car dealerships and was based on introducing five artificial standard alternatives ($SA = \{A, B, C, D, E\}$) together with service $X$ to be evaluated. The classification of $X$ varies according to the position of $X$ in the ranking outputted from Electre III algorithm.

Later, Costa et al. [2007] reviewed the proposal described in Freitas and Costa [1998] by changing the Electre III algorithm to the more pessimistic or exigent procedure of ELECTRE TRI Mousseau and Slowinski [1998]. The main advantage of using Electre TRI instead of Electre III was the reduction of computational efforts when compared with Electre III, since there was no need to compare alternatives among them. As described in Costa et al. [2007], the Electre TRI was applied to evaluate the customers' satisfaction regarding the services provided by a candy store.

Lupo [2015] applied the ELECTRE III for evaluating the quality of services provided by three international airports located in Sicilia, Italy. In this modeling, AHP is used for eliciting the weights of the criteria, and the average perceived quality on the $k^{th}$ service attribute was estimated by aggregating quality scores of related service attributes via the arithmetic mean operator mean.

Barbosa et al. [2018a] evaluated the performance of electricity distribution utilities with the use of a single global index by means of the Analytic Hierarchy Process (AHP) and Preference Ranking Organization Method for Enrichment

Evaluations (PROMETHEE) methods. The proposed approach allows the ranking of service quality according to three dimensions: supply continuity, voltage conformity, and customer satisfaction. AHP was used to set the weights for the criteria, and PROMETHEE showed the results in the form of a ranking. This made it easier for regulators to judge the performance of the distributors, which improved the quality of the services that utilities provided. La Fata et al. [2019] applied a similar modeling to that described in (Lupo 2015a), in order to rank the quality of health services. In this work, AHP was only used to elicit weights, not rank them. We classify these works as having applied a non-compensatory ranking approach.

In a different area, Sosyal et al. [2017] used the Fuzzy Analytic Hierarchy Process (FAHP) to figure out how important e-service quality factors were by surveying the people who made the decisions. The Fuzzy PROMETHEE approach evaluated the e-service quality of four websites belonging to Turkish aviation businesses, using data from customer surveys.

In the research described in Tuzkaya et al. [2019], PROMETHEE was applied to a real-life case study of an Istanbul public hospital; the proposal was used to evaluate service quality based on patient feedback. The evaluation provides thorough information on the impact of criteria and identifies possibilities to improve service quality through patient feedback analysis.

Costa and Duarte [2019] proposed a model for evaluating the Quality of Services (QOS) of libraries under the viewpoint of multiple evaluators that take into account multiple criteria. In that study, ELECTRE TRI ME was applied to classify the services provided by a library based on the perceptions of 72 library users. The ELECTRE TRI ME was also used in Medeiros Da Rocha et al. [2022], that proposed a model for evaluating the Quality of Services (QOS) of the international airport of Rio de Janeiro (SATA code: GIG) under the perspective of perceptions of a sample composed of 1935 customers that evaluate the services provided by the International Airport Antonio Carlos Jobim (SBGL), located in Rio de Janeiro, Brazil.

The newest study on using an outranking (PROMETHEE) method to judge the quality of services Yao et al. [2023] is based on a poll of 210 experienced users of delivery platforms. Four food delivery platforms at Twain were ranked using an overall score from modified PROMETHEE by comparing their actual performance to their aspiration level.

One can observe that most of the modelling mentioned in this section pre-processes the data by using a compensatory method, which should introduce noise in the modelling: to input data computed through a compensatory procedure into a no-compensatory sorting algorithm. The exceptions are Costa and Duarte [2019] and Medeiros Da Rocha et al. [2022], which applied a full non-compensatory approach for classifying the quality of services provided, respectively, by a library and an airport terminal of passengers.

## 3 Compensatory vs. Non-compensatory Multi-criteria Aggregation Methods

In this section, we highlight the main differences between compensatory and non-compensatory multi-criteria methods for aggregating preferences. For a deeper discussion about this subject, we suggest reading Costa e pessoa (2023)

### 3.1 Compensatory methods

As it appears in Equation 1, compensatory aggregation is usually based on computing the total utility of an object $x$ scalar, which value is equal to the weighted sum of the utilities of the alternative $(x)$ considering an entire set of criteria composed of $n$ criteria.

$$U(x) = \sum_{j=1}^{m} k_j * u(x_j) \qquad (1)$$

where:

- $n$ is the number of criteria used in the modeling
- $k_j$ is the constant of scale (sometimes called as weight) of the $j^{th}$ criterion.

The multi-criteria methods that implement additive approaches are usually classified as based on MAUT, which is presented in the seminal works of **?** and Keeney and Raiffa [1993]. Nowadays, the most used MAUT-based methods are TOPSIS [Hwang and Yoon, 1981], AHP [Saaty, 1980], and FiTradeoff [De Almeida et al., 2016]. Because MAUT modeling is additive, it can show compensatory behavior. This means that if $x$ has a low performance in one criterion, a high performance under criterion $j$ should make up for it.

Notice that, when applied in ranking problems, the methods based on MAUT first compute for each alternative the value of the utility function and then rank the alternatives according to the values of their utility function. In other words, when applied in ranking problems, the methods based on MAUT compute for each alternative the values of the utility function and, after such computing, rank the alternatives.

## 3.2 Non-compensatory methods

On the other hand, there are the non-compensatory methods, which do not compute an overall utility function as the one that appears in 1. These methods are based on comparing the performance of the alternatives, looking to discover to which degree an alternative covers or outranks the other ones—that is because these methods are usually referred to as outranking ones. $S(a,b)$ means that $a$ outranks $b$.

Table 2 shows a summary of the most known multi-criteria outranking methods. Their main feature is their capacity to avoid compensatory effects that should be undesirable in some decision situations. The Equation 2 shows a simplified version of the credibility function $\sigma(a,b)$ that is used in ELECTRE III to figure out the degree of credibility with the statement $S(a,b)$ ($a$ is more credible than $b$).

$$\sigma(a,b) = \left[\frac{1}{\sum_{j=1}^{n} w_j}\right] * \sum_{j=1}^{n} w_j * c_j(a,b) \quad (2)$$

$$c_j(a,b) = \begin{cases} 1 & \iff u_j(a) \geq u_j(b) \\ 0 & \text{otherwise} \end{cases}$$

Where:

- $c_j(a,b)$ is a local concordance degree; that means the concordance in the $j^{th}$ criterion with the assertive "$a$ is at least as good as $b$ in the criterion $j$.
- $n$ is the number of criteria used in the modeling.
- $w_j$ is the constant of scale (sometimes called a weight) of the $j^{th}$ criterion.

Table 2: Summary of outranking methods

| Type | Method | Reference | Number of evaluators |
|---|---|---|---|
| Choice | Electre I | Roy [1968] | 1 |
| | Electre IS | Roy and Skalka [1984] | 1 |
| | Electre I ME | Costa et al. | $n \geq 1$ |
| Ranking | Electre II | Roy and Bertier [1971] | 1 |
| | Electre III | Roy [1978] | 1 |
| | Promethee II | Brans and Mareschal [2005] | 1 |
| Sorting | Electre TRI | Mousseau et al. [2000] | 1 |
| | Electre TRI-C | Almeida-Dias et al. [2010] | 1 |
| | Electre TRI-nC | Almeida-Dias et al. [2012] | 1 |
| | Electre TRI ME | Costa et al. [2020] | $n \geq 1$ |

Where:

- $c_j(a,b)$ is a local concordance degree; that means the concordance in the $j^{th}$ criterion with the assertive "$a$ is at least as good as $b$ in the criterion $j$.

- $n$ is the number of criteria used in the modeling.
- $w_j$ is the constant of scale (sometimes called a weight) of the $j^{th}$ criterion.

# 4 Fundamentals of the ELECTRE TRI ME

As the modelling in this paper is an evolution of the ELECTRE TRI ME (see algorithm B in Appendix B ) to face the problem of LoS evaluation, this section shows a summary of this method.

To illustrate some features of the data to be inputted into ELECTRE TRI ME, it follows the description of a hypothetical "toy example" of sorting the quality of the services in a hotel. For simplicity, in this situation, without loss of generality, a set of three customers $\{E = e_1, e_2, e_3, \}$ evaluated the services provided by a hotel $X$ according to a five-point scale of satisfaction. Table 3 summarizes this situation.

Table 3: Hypothetical situation in which three evaluators evaluate the service X using their own criteria set and boundaries

|  | Evaluators | | | | | | | |
| --- | --- | --- | --- | --- | --- | --- | --- | --- |
|  | $e_1$ | | $e_2$ | | | $e_3$ | | |
|  | Check_in | Acessibility | Room services | Overall cleaness | Gymnasium | Check-in | Restaurant | Room services |
| Criteria weight | 10 | 8 | 5 | 10 | 6 | 8 | 7 | 10 |
| Boundary 1 | 4.5 | 4.5 | 4.5 | 4.5 | 4.5 | 4.5 | 4.5 | 4.5 |
| Boundary 2 | 3.5 | 3.5 | 3.5 | 3.5 | 3.5 | 3.5 | 3.5 | 3.5 |
| Boundary 3 | 2.5 | 2.5 | 2.5 | 2.5 | 2.5 | 2.5 | 2.5 | 2.5 |
| Boundary 4 | 1.5 | 1.5 | 1.5 | 1.5 | 1.5 | 1.5 | 1.5 | 1.5 |
| Evaluation of **X** | 3 | 2 | 2 | 1 | 2 | 3 | 5 | 5 |

- The set $A$ of alternatives is unitary, that is, it has only one alternative: $x \in A$.
- Each evaluator has its own criteria set and criteria weights, so that:

    $Fe_1 = Check - in, Acessibility$
    $We_1 = 10, 8$
    $Fe_2 = Room\ services, Breakfeast, Gymnasium$
    $We_2 = 5, 10, 6$
    $Fe_3 = Check - in, Restaurant, Roomservices$
    $We_3 = 8, 7, 10$

- For all criteria
$$C = \{Very\ good, Good, Middle, Poor, Very\ poor\}$$
    Or, using a numerical scale:
$$C = \{5, 4, 3, 2, 1\}$$
    Which implies in: :
$$B = \begin{bmatrix} 4.5 & 4.5 & 4.5 & 4.5 & 4.5 & 4.5 & 4.5 & 4.5 \\ 3.5 & 3.5 & 3.5 & 3.5 & 3.5 & 3.5 & 3.5 & 3.5 \\ 2.5 & 2.5 & 2.5 & 2.5 & 2.5 & 2.5 & 2.5 & 2.5 \\ 1.5 & 1.5 & 1.5 & 1.5 & 1.5 & 1.5 & 1.5 & 1.5 \end{bmatrix}$$

- The performance of $x \in A$ under the criteria set $F = Fe_1 \cup Fe_2 \cup Fe_3$ is:

Table 4: Credibility degrees for the toy example

| Credibility relationship | Computation | Credibility degree |
|---|---|---|
| $\sigma(X, Very\ good)$ | $\frac{0+0+0+0+0+0+7+10}{10+8+5+10+6+8+7+10}$ | $\frac{17}{64}$ |
| $\sigma(X, Good)$ | $\frac{0+0+0+0+0+0+7+10}{10+8+5+10+6+8+7+10}$ | $\frac{17}{64}$ |
| $\sigma(X, Middle) =$ | $\frac{10+0+0+0+0+8+7+10}{10+8+5+10+6+8+7+10} =$ | $\frac{35}{64}$ |
| $\sigma(X, Poor)$ | $\frac{7+10}{10+8+5+0+6+8+7+10}$ | $\frac{54}{64}$ |
| $\sigma(X, Very\ poor)$ | $\frac{10+8+5+10+6+8+7+10}{10+8+5+10+6+8+7+10}$ | $\frac{64}{64}$ |

$$G = \begin{bmatrix} 4.5 & 4.5 & 4.5| & 4.5 & 4.5 & 4.5 & 4.5 & 4.5 \\ 3.5 & 3.5 & 3.5 & 3.5 & 3.5 & 3.5 & 3.5 & 3.5 \\ 2.5 & 2.5 & 2.5 & 2.5 & 2.5 & 2.5 & 2.5 & 2.5 \\ 1.5 & 1.5 & 1.5 & 1.5 & 1.5 & 1.5 & 1.5 & 1.5 \end{bmatrix}$$

If we have applied the ELECTRE ME algorithm that appears in B with a cut level $\lambda = 0.70$, it will results in the values of credibility degree shown in Table 4 and in classifying the LoS of **X** as "Poor" level.

## 5 Proposition

The proposal is based on adapting Algorithm Appendix B (ELECTRE TRI ME) described in Appendix B through substituting step 3 by the following ones:

[**Step 3:** ] Build matrix ($SM$) to evaluate sorting sensibility based on the cut-level ($\lambda$).

[**Step 4:** ] Group airports according the sensibility matrix.

[**Step 5:** ] Propose customized actions.

Figure 1 shows the steps flow of the proposal. The data to be collected in the first step are:

- Set $A \leftarrow \{a_1, a_2, \ldots a_m\}$ of $m$ instances to be evaluated.
- Set $E \leftarrow \{e_1, e_2, \ldots e_n\}$, composed by $n$ evaluators.
- Family $F$ of criteria. As each evaluator can have its own criteria set, them $F \leftarrow Fe_1 \cup Fe_2 \cup \ldots \cup Fe_n$, where:

    $Fe_1 \leftarrow \{k_{1e_1}, k_{2e_1}, \ldots k_{ve_1}\}$ is the subset composed by the $v$ criteria adopted by the evaluator $e_1$.
    $Fe_2 \leftarrow \{k_{1e_2}, k_{2e_2}, \ldots k_{xe_2}\}$ is the subset composed by the $x$ criteria adopted by the evaluator $e_2$.
    $Fe_j \leftarrow \{k_{1e_j}, k_{2e_j}, \ldots k_{e_2}\}$ is the subset composed by the $y$ criteria adopted by the evaluator $e_j$.
    $Fe_n \leftarrow \{k_{1e_n}, k_{2e_n}, \ldots k_{xe_n}\}$ is the subset composed by the $z$ criteria adopted by the evaluator $e_n$.
    $F \leftarrow Fe_1 \cup Fe_2 \cup \ldots \cup Fe_n$

- Weights (or constant of scales) of criteria $W \leftarrow We_1 \cup We_2 \cup \ldots \cup We_n$, where:

    $We_1 \leftarrow \{w_{1e_1}, w_{2e_1}, \ldots w_{ve_1}\}$ is the subset composed by the weights of $Fe_1$.
    $We_2 \leftarrow \{w_{1e_2}, w_{2e_2}, \ldots w_{xe_2}\}$ is the subset composed by the weights of $Fe_2$.
    $We_j \leftarrow \{w_{1e_j}, w_{2e_j}, \ldots w_{e_2}\}$ is the subset composed by the weights of $Fe_j$.
    $We_n \leftarrow \{w_{1e_n}, w_{2e_n}, \ldots w_{xe_n}\}$ is the subset composed by the weights of $Fe_n$.

- Performance of alternatives
    $G \leftarrow Ge_1(a_1) \cup \ldots Ge_j(a_i) \cup \ldots \cup Ge_n(a_m)$, where:

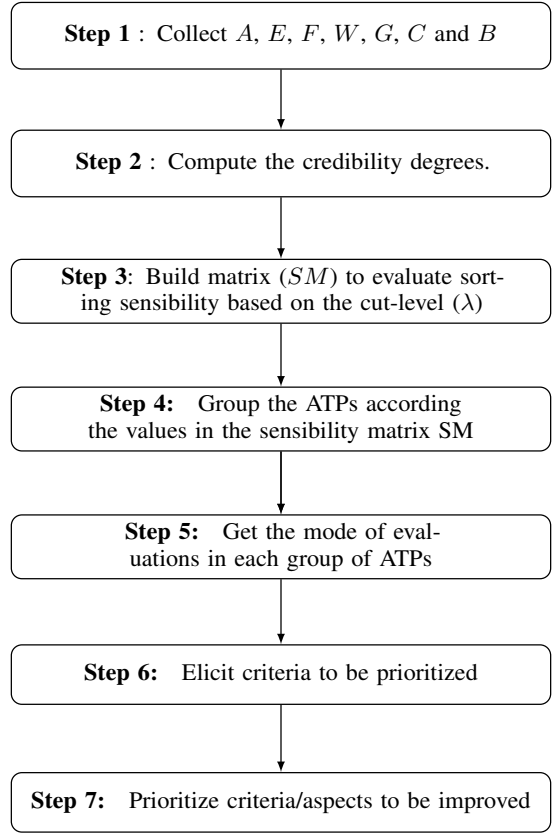

Figure 1: Graphical representation of the research flow

$Ge_1(a_1) \leftarrow \{g_{1e_1}(a_1), g_{2e_1}(a_1), \ldots g_{ve_1}(a_1)\}$ is the subset composed by the performance of alternative $a_1$ under the perspective of the evaluator $e_1$ and the set of criteria $Fe_1$. For example, $g_{2e1}(a_i)$ means the performance of alternative $a_i$ under criterion 2 and under the viewpoint of the evaluator $e_1$.

$ge_j(a_i) \leftarrow \{g_{1e_j}(a_i), g_{2e_j}(a_i), \ldots g_{|Fe_j|e_j}(a_i)\}$ is the subset composed by the performance of alternative $a_i$ under the perspective of the evaluator $e_j$ and the set of criteria $Fe_j$.

. . .

$ge_n(a_m) \leftarrow \{g_{1e_n}(a_m), \ldots, g_{|Fe_j|e_n}(a_m)\}$ is the subset composed by the performance of alternative $a_m$ under the perspective of the evaluator $e_n$ and the set of criteria $Fe_n$.

- Set of categories $C \leftarrow \{C_1, C_2, \ldots C_k\}$ in which the instances will be sorted.

  $C$ is ranked from the best to the worst, i.e. $C_1$ is the best category and $C_k$ is the worst one. The number of categories is the same for all $e \in E$ and all $Fe \in F$.

- Set of categories' boundaries $B$

  Set of boudaries $B$, notice that $|B|$ is equal to $|C| - 1$, and that $b_{pfe}$ is the boundary of the $p^{th}$ category, in the $f^{th}$ criterion for the $e^{th}$ evaluator. For example, $b_{123}$ is the boundary of category 1, in criterion 2 for evaluator 3.

  $B \leftarrow Be_1 \cup Be_2 \cup \ldots \cup Be_n.$

# 6 Results from applying the proposition for evaluating LoS in Airport Terminal of Passengers

In this section, we describe the steps of instancing the proposal in case of evaluation of fifteen Airport Terminal of Passengers (ATP) located in Brazil.

Table 5: Sample of airport passenger terminals (ATP)

| IATA | Denomination | City | Sample size |
|---|---|---|---|
| BSB | Presidente Juscelino Kubitschek | Brasília | 1279 |
| BEL | Val de Cans | Belém | 1275 |
| CNF | Tancredo Neves | Confins | 1141 |
| CWB | Afonso Pena | Curitiba | 954 |
| FLN | Hercílio Luz | Florianópolis | 1110 |
| FOR | Pinto Martins | Fortaleza | 1192 |
| GIG | Galeão, Internacional do Rio de Janeiro | Rio de Janeiro | 1976 |
| GRU | Guarulhos, Internacional de São Paulo | Guarulhos | 1989 |
| MAO | Eduardo Gomes | Manaus | 1126 |
| MCZ | Zumbi dos Palmares | Maceió | 985 |
| NAT | Governador Aluízio Alves | Natal | 854 |
| POA | Salgado Filho | Porto Alegre | 1431 |
| REC | Guararapes, Gilberto Freyre | Recife | 1132 |
| SSA | Deputado Luís Eduardo Magalhães | Salvador | 1323 |
| VCP | Viracopos | Campinas | 1473 |

## 6.1 Collect the data

The information of this study is based on the general report on operational performance indicators in airports, released by SAC (SAC is an acronimous in Portuguese that means the National Civil Aviation Department, which is located at Brazil). The report is the result of an *in loco* survey performed by a independent consulting firm hired by SAC, which collected data through the application of standardized questionnaires in face-to-face interviews with passengers traveling through the analyzed airports. Each interviewed passenger assigned grades from 4, the lowest satisfaction level possible, up to 4, that means the highest level of satisfaction, according to 38 indicators. All the primary data used during the study were provided by a third party. Direct request for these materials may be made to the provider: the Brazilian National Civil Aviation Department (SAC).

In the specific case of Brazil, due to its continental dimensions, civil aviation plays a strategic role for the development of the country, both by the circulation of people and goods it provides, as well as by the generation of jobs and financial movement. In addition, another factor worth mentioning is the process of granting airport infrastructure to private sector, promoted by the Brazilian government through auctions and concession contracts. The airport concession has the main objective of attracting investments to expand and improve the Brazilian airport infrastructure, and thus, promote improvements in the service to air transport users.

As a result of these activities, several factors associated with the Brazilian airport infrastructure and issues related to the state of the airport's operational performance, as well as its form of assessment from the perspective of passengers, became a topic of discussion. The uncertainty regarding the capacity of the existing infrastructure to adequately supply demand was placed on the agenda, as expected by users. It has thus become relevant to identify methodologies that assess the service level of airport passenger terminals, in order to enable the planning of future availability, consistent with the quality standards desired by society.

### 6.1.1 Alternatives or airports to be sorted

Nowadays, the survey involves 20 airports, with the largest movement in the country, responsible for almost 90% of the total passengers transported by Brazilian regular aviation. However, five of them operate only domestic flights and, therefore, will not be included in the application of this study, since they are not evaluated under the entire set of criteria.

Table 5 shows the set $A$ composed by 15 airport terminals of passenger (ATP) located in Brazil, all of them operating international flights. The first the columns present the ICAO and IATA codes, respectively, and, the third one inform the Brazilian city in which the airport is localized.

### 6.1.2 Evaluators for each airport passenger terminal

In order to reinforce the final customer of the service as the main actor in the process, it is proposed that passengers be the evaluators. Different profiles are also considered: departing, transferring and arriving passengers. In this way, all user profiles are taken into account in the decision process.

As a consequence, the evaluators are the members of a set composed by $n = 19,240$ users that volunteered to answer a questionnaire applied face-to-face. The last column of Table 5 shows the number of passengers interviewed at each of the airports involved in the survey.

### 6.1.3 Criteria

The operational performance indicators adopted in the survey conducted by SAC were proposed by the Brazilian Operational Performance Committee of National Airport Authorities Commission (CONAERO) and composed by public entities involved in the processing of aircraft, passengers and goods at Brazilian airports. The purpose of the indicators is to assess the passenger experience in several items of infrastructure, service provision, as well as to monitor the performance of different airport processes such as check-in, security inspection, baggage claim, among others. Table 6 shows the starting set of criteria (operational performance indicators) presented to the passengers, so that they could choose their own subset to perfom their evaluations.

For example, suppose an evatuor $e_j$ who had choose to evaluate the AT using the criteria "Courtesy and helpfulness of security staff", C6 "Flight information display screens (FIDS)", and, "Charging stations (battery recharge facilities)". For this passenger the criteria set would be:

$$Fe_j = \{C4, C6, C7\}$$

### 6.1.4 Criteria's weights

The form applied in the survey did not collected perceptions about the immportance of criteria. Therefore, in agreement with the Operational Performance Committee/CONAERO, all operational performance indicators were considered equally important in their contribution to the development of airport management and to increase the capacity and efficiency of the infrastructure of Brazilian airports. In this sense, equality of the criteria was adopted in this proposal, so that:

$$w_j = 1, \ \forall \, wj \in W$$

### 6.1.5 Performance of airports

A standard questionnaire was used to conduct interviews with passengers, which took place in the departure and arrival lounges of airport terminals. Each passenger evaluated the criteria wished and felt comfortable with (among that are part of the SC), based on their experience at the airport. In other words, each evaluator had its own set of criteria. For this, evaluators assigned discrete scores from 1 to 5 for each criterion, where 1 was the worst possible score and 5 was the best possible score, according the options in Table 7.

The sample distribution of the interviews was performed according to the flow of passengers, considering a maximum margin error of 5% with a 95% confidence interval. This stratification aims to guarantee the suitability of the sample, with the collections carried out during the hours of greatest flow of passengers at the airports, defined together with the airport operators, in order to obtain passenger opinion when the airport presents the greatest concentration of activities in operation.

### 6.1.6 Set of categories $C$

The categories need to be completely aligned with the main focus of the survey and the scale used to evaluate the perceptions about the service performance. Therefore, as it appears in Table 8, the categories were based on the options that respondents could choose in the questionnaire.

### 6.1.7 Boundaries

The boundaries are the "floor" of categories $C_1 \ldots C_4$. Hence, based on the the Tables 7 and 8, we defined the boundaries that appear in Table 9. Observe that the lowest category $C_5$ does not need to have a lower boundary defined.

## 6.2 Compute the credibility degree

Table 10 shows the values gotten for the credibility degree, obtained in **Step 2** of Algorithm B. As an example, from this table there is a credibility level equal to 53,4% with classifying GIG as $C_1$, and equal to 83,6% with classifyng this airport at least as $C_2$, once:

- $\sigma(GIG, C_1) = 0.534$, and,
- $\sigma(GIG, C_2) = 0.836$

Table 6: Initial set of criteria (SC)

| Code | Criteria |
| --- | --- |
| Cr1 | Easy embarkation/disembarkation (curbside) |
| Cr2 | Waiting time at security inspection |
| Cr3 | Thoroughness/efficiency of security inspection |
| Cr4 | Courtesy and helpfulness of security staff |
| Cr5 | Directions and signage (Ease of finding your way through airport) |
| Cr6 | Flight information display screens (FIDS) |
| Cr7 | Charging stations (battery recharge facilities) |
| Cr8 | Internet access and Wi-fi |
| Cr9 | Availability of washrooms/toilets |
| Cr10 | Cleanliness of washrooms/toilet |
| Cr11 | Availability of seats (Departure lounge) |
| Cr12 | Feeling of being safe and secure (Public areas) |
| Cr13 | Cleanliness |
| Cr14 | Thermal comfort |
| Cr15 | Acoustic comfort |
| Cr16 | Quality of information on baggage claim conveyor display screens |
| Cr17 | Quality of parking facilities |
| Cr18 | Availability of parking spaces |
| Cr19 | Value for money of parking facilities (prices) |
| Cr20 | Restaurant/Eating facilities (availability and quality) |
| Cr21 | Value for money of restaurant/eating facilities (prices) |
| Cr22 | Availability of bank/ATM facilities/money changers |
| Cr23 | Shopping facilities (availability and quality) |
| Cr24 | Value for money of shopping facilities (prices) |
| Cr25 | Self-check-in facilities waiting time |
| Cr26 | Check-in waiting time |
| Cr27 | Courtesy and helpfulness of check-in staff |
| Cr28 | Quality of information provided by airline |
| Cr29 | Baggage delivery service (speed) |
| Cr30 | Integrity of baggage delivered |
| Cr31 | Waiting time at passport/personal ID inspection |
| Cr32 | Courtesy and helpfulness of inspection staff |
| Cr33 | Waiting time for immigration processing |
| Cr34 | Courtesy of immigration bureau staff |
| Cr35 | Waiting time for customs inspection |
| Cr36 | Courtesy of customs staff |
| Cr37 | Ground transportation to/from airport |

### 6.3 Build matrix $SM$ to evaluate sorting sensibility based on the cut-level ($\lambda$)

Table 11 shows the variations in the sorting of the ATP according the credibility cut-levels in the interval $[0.50\ldots1.00]$.

### 6.4 Group the ATPs according the values in the sensibility matrix SM

A visual examination of this table suggests the aggregation of ATP in the groups displayed in Table 12. The third column of this shows the position of the cluster in a ranking from the best performance to the worst. The last column displays a category code created for referring to the group.

### 6.5 Get the mode of evaluations in each group of ATPs

To provide additional information to support the discussion, we put on Table 13 the values of the mode metric in each criterion regarding the evaluations received by the ATPs in the same category. For example: the mode of the evaluations received by ATPs in Group D regarding criterion $Cr5$ was 4.

Table 7: Scale for evaluating the performance

| Verbal value | Numerical Value |
|---|---|
| Very Bad | 1 |
| Bad | 2 |
| Fair | 3 |
| Good | 4 |
| Very Good | 5 |

Table 8: Categories

| Code | Verbal value |
|---|---|
| $C_1$ | Very Good |
| $C_2$ | Good |
| $C_3$ | Fair |
| $C_4$ | Bad |
| $C_5$ | Very Bad |

### 6.6 Elicit criteria to be improved

As a whole, in most criteria the ATPs had a performance perceived at least as "Good" or "very good". The exceptions are the the criteria "Cr24: Value for money of shopping facilities (prices)" and "Cr21: Value for money of restaurant/eating facilities (prices)", that were evaluated as 3, which means a "Regular" level of service, in all ATP's groups.

We also observe that, for Groups A. B, e C, the mode of the evaluations in the question "Overall Performance" was "Very Good" (the highest level in c the scale), which means that ATPs in Group D should be estimulated to improve their performances mainly in those criteria that they have a performance lower than the mode of the upper categories.

By analyzing the results that appears in Table 11 and Table 12, we observe that ATPs classified in:

- Group A has the best performances, maintaining mainly service level "Very Good" in 32/37 criteria. Therefore, beyond adopting actions prioritizing the improvement of their performance in criteria Cr21 (Value for money of restaurant/eating facilities (prices)) and Cr24 (Value for money of shopping facilities (prices)), the ATP in this category should also improve their performance in the following criteria:
    - Cr19: value for money of parking facilities (prices)
    - Cr20: Restaurant/Eating facilities (availability and quality)
    - Cr22: Availability of bank/ATM facilities/money changers
- Groups B and C, respectively, have the same mode in all criteria. The gap in the mode in Group A, is really slight and it occurs in the criteria Cr19 (Value for money of parking facilities (prices)), Cr20 (Restaurant/Eating facilities (availability and quality)), and Cr22 (Availability of bank/ATM facilities/money changers). Hence, ATPs in these groups B and C should make actions to improve their performance in the following criteria:
    - Cr17: Quality of parking facilities
    - Cr21: Value for money of restaurant/eating facilities (prices)
    - Cr23: Shopping facilities (availability and quality)
- Group D have the lowest performances among the ATPs studied. Tthe ATP in $D$ should emphasize the prioritization of actions to improve their performance in these criteria.
    - Cr5: Directions and signage (Easy of finding your way through airport)
    - Cr6: Flight information display screens (FIDS)

Table 9: Boundaries of the categories

| Code | Boundary values | Description |
|---|---|---|
| $B_1$ | $b_{1je} = 4,5, \forall f_j \in F$ | Lower boundary of $C_1$ |
| $B_2$ | $b_{2je} = 3,5, \forall f_j \in F$ | Lower boundary of $C_2$ |
| $B_3$ | $b_{3je} = 2,5, \forall f_j \in F$ | Lower boundary of $C_3$ |
| $B_4$ | $b_{4je} = 1,5, \forall f_j \in F$ | Lower boundary of $C_4$ |

Table 10: Credibility degree

| IATA | $B_1$ | $B_2$ | $B_3$ | $B_4$ |
|---|---|---|---|---|
| BSB | 0.501 | 0.842 | 0.953 | 0.983 |
| BEL | 0.461 | 0.792 | 0.932 | 0.977 |
| CNF | 0.584 | 0.907 | 0.973 | 0.990 |
| CWB | 0.711 | 0.920 | 0.976 | 0.991 |
| GIG | 0.534 | 0.836 | 0.948 | 0.981 |
| GRU | 0.494 | 0.820 | 0.934 | 0.971 |
| FLN | 0.515 | 0.820 | 0.947 | 0.983 |
| FOR | 0.523 | 0.801 | 0.930 | 0.975 |
| MCZ | 0.615 | 0.891 | 0.968 | 0.989 |
| MAO | 0.524 | 0.840 | 0.948 | 0.981 |
| NAT | 0.569 | 0.855 | 0.948 | 0.979 |
| POA | 0.520 | 0.807 | 0.934 | 0.977 |
| REC | 0.473 | 0.805 | 0.941 | 0.977 |
| SSA | 0.416 | 0.751 | 0.910 | 0.966 |
| VCP | 0.652 | 0.892 | 0.967 | 0.989 |

Table 11: Sensibility of sorting according the cut-levels ($\lambda$) applied to credibility degree

| IATA | 0.05 | 0.10 | 0.15 | 0.20 | 0.25 | 0.30 | 0.35 | 0.40 | 0.45 | 0.50 | 0.55 | 0.60 | 0.65 | 0.70 | 0.75 | 0.80 | 0.85 | 0.90 | 0.95 | 1.00 |
|---|---|---|---|---|---|---|---|---|---|---|---|---|---|---|---|---|---|---|---|---|
| BEL | $C_1$ | $C_1$ | $C_1$ | $C_1$ | $C_1$ | $C_1$ | $C_1$ | $C_1$ | $C_1$ | $C_2$ | $C_2$ | $C_2$ | $C_2$ | $C_2$ | $C_2$ | $C_3$ | $C_3$ | $C_3$ | $C_4$ | $C_5$ |
| BSB | $C_1$ | $C_1$ | $C_1$ | $C_1$ | $C_1$ | $C_1$ | $C_1$ | $C_1$ | $C_1$ | $C_1$ | $C_2$ | $C_2$ | $C_2$ | $C_2$ | $C_2$ | $C_3$ | $C_3$ | $C_3$ | $C_3$ | $C_5$ |
| CNF | $C_1$ | $C_1$ | $C_1$ | $C_1$ | $C_1$ | $C_1$ | $C_1$ | $C_1$ | $C_1$ | $C_1$ | $C_2$ | $C_2$ | $C_2$ | $C_2$ | $C_2$ | $C_2$ | $C_2$ | $C_2$ | $C_3$ | $C_5$ |
| CWB | $C_1$ | $C_1$ | $C_1$ | $C_1$ | $C_1$ | $C_1$ | $C_1$ | $C_1$ | $C_1$ | $C_1$ | $C_1$ | $C_1$ | $C_1$ | $C_2$ | $C_2$ | $C_2$ | $C_2$ | $C_3$ | $C_3$ | $C_5$ |
| FLN | $C_1$ | $C_1$ | $C_1$ | $C_1$ | $C_1$ | $C_1$ | $C_1$ | $C_1$ | $C_1$ | $C_1$ | $C_2$ | $C_2$ | $C_2$ | $C_2$ | $C_2$ | $C_2$ | $C_3$ | $C_3$ | $C_4$ | $C_5$ |
| FOR | $C_1$ | $C_1$ | $C_1$ | $C_1$ | $C_1$ | $C_1$ | $C_1$ | $C_1$ | $C_1$ | $C_2$ | $C_2$ | $C_2$ | $C_2$ | $C_2$ | $C_2$ | $C_2$ | $C_3$ | $C_3$ | $C_4$ | $C_5$ |
| GIG | $C_1$ | $C_1$ | $C_1$ | $C_1$ | $C_1$ | $C_1$ | $C_1$ | $C_1$ | $C_1$ | $C_2$ | $C_2$ | $C_2$ | $C_2$ | $C_2$ | $C_2$ | $C_2$ | $C_3$ | $C_3$ | $C_4$ | $C_5$ |
| GRU | $C_1$ | $C_1$ | $C_1$ | $C_1$ | $C_1$ | $C_1$ | $C_1$ | $C_1$ | $C_1$ | $C_2$ | $C_2$ | $C_2$ | $C_2$ | $C_2$ | $C_2$ | $C_2$ | $C_3$ | $C_3$ | $C_4$ | $C_5$ |
| MAO | $C_1$ | $C_1$ | $C_1$ | $C_1$ | $C_1$ | $C_1$ | $C_1$ | $C_1$ | $C_1$ | $C_1$ | $C_2$ | $C_2$ | $C_2$ | $C_2$ | $C_2$ | $C_2$ | $C_3$ | $C_3$ | $C_4$ | $C_5$ |
| MCZ | $C_1$ | $C_1$ | $C_1$ | $C_1$ | $C_1$ | $C_1$ | $C_1$ | $C_1$ | $C_1$ | $C_1$ | $C_1$ | $C_1$ | $C_2$ | $C_2$ | $C_2$ | $C_2$ | $C_2$ | $C_3$ | $C_3$ | $C_5$ |
| NAT | $C_1$ | $C_1$ | $C_1$ | $C_1$ | $C_1$ | $C_1$ | $C_1$ | $C_1$ | $C_1$ | $C_1$ | $C_2$ | $C_2$ | $C_2$ | $C_2$ | $C_2$ | $C_2$ | $C_2$ | $C_3$ | $C_4$ | $C_5$ |
| POA | $C_1$ | $C_1$ | $C_1$ | $C_1$ | $C_1$ | $C_1$ | $C_1$ | $C_1$ | $C_1$ | $C_2$ | $C_2$ | $C_2$ | $C_2$ | $C_2$ | $C_2$ | $C_2$ | $C_3$ | $C_3$ | $C_4$ | $C_5$ |
| REC | $C_1$ | $C_1$ | $C_1$ | $C_1$ | $C_1$ | $C_1$ | $C_1$ | $C_1$ | $C_2$ | $C_2$ | $C_2$ | $C_2$ | $C_2$ | $C_2$ | $C_2$ | $C_3$ | $C_3$ | $C_3$ | $C_4$ | $C_5$ |
| SSA | $C_1$ | $C_1$ | $C_1$ | $C_1$ | $C_1$ | $C_1$ | $C_1$ | $C_1$ | $C_2$ | $C_2$ | $C_2$ | $C_2$ | $C_2$ | $C_2$ | $C_2$ | $C_3$ | $C_3$ | $C_3$ | $C_4$ | $C_5$ |
| VCP | $C_1$ | $C_1$ | $C_1$ | $C_1$ | $C_1$ | $C_1$ | $C_1$ | $C_1$ | $C_1$ | $C_1$ | $C_1$ | $C_1$ | $C_1$ | $C_2$ | $C_2$ | $C_2$ | $C_2$ | $C_3$ | $C_3$ | $C_5$ |

Table 12: ATPs organized into $K = 4$ groups

| Group | ATP | Ranking |
|---|---|---|
| A | VCP, and, CWB | $1^{st}$ |
| B | CNF, MCZ, and, NAT | $2^{nd}$ |
| C | BSB, GIG, GRU, FOR, POA, FLN, and, MAO | $3^{rd}$ |
| D | REC, SSA, and, BEL | $4^{th}$ |

– Cr8: Internet access and Wi-fi
– Cr9: Availability of washrooms/toilets
– Cr10: Cleanliness of washrooms/toilet
– Cr13: Cleanliness
– Cr15: Acoustic comfort
– Cr18: Availability of parking spaces

Beyond rising their performance in the mentioned criteria, the ATPs in Group D should also improve their performances in the following criteria:

– Cr17: Quality of parking facilities
– Cr19: value for money of parking facilities (prices)

Table 13: Mode of the evaluations for each category

| Criterion | Group A | B | C | D |
| --- | --- | --- | --- | --- |
| Cr1 | 5 | 5 | 5 | 5 |
| Cr2 | 5 | 5 | 5 | 5 |
| Cr3 | 5 | 5 | 5 | 5 |
| Cr4 | 5 | 5 | 5 | 5 |
| Cr5 | 5 | 5 | 5 | 4 |
| Cr6 | 5 | 5 | 5 | 4 |
| Cr7 | 5 | 5 | 5 | 5 |
| Cr8 | 5 | 5 | 5 | 4 |
| Cr9 | 5 | 5 | 5 | 4 |
| Cr10 | 5 | 5 | 5 | 4 |
| Cr11 | 5 | 5 | 5 | 5 |
| Cr12 | 5 | 5 | 5 | 5 |
| Cr13 | 5 | 5 | 5 | 4 |
| Cr14 | 5 | 5 | 5 | 5 |
| Cr15 | 5 | 5 | 5 | 4 |
| Cr16 | 5 | 5 | 5 | 5 |
| Cr17 | 5 | 4 | 4 | 4 |
| Cr18 | 5 | 5 | 5 | 4 |
| Cr19 | 4 | 3 | 3 | 3 |
| Cr20 | 4 | 4 | 4 | 4 |
| Cr21 | 3 | 3 | 3 | 3 |
| Cr22 | 4 | 4 | 4 | 4 |
| Cr23 | 5 | 4 | 4 | 4 |
| Cr24 | 3 | 3 | 3 | 3 |
| Cr25 | 5 | 5 | 5 | 5 |
| Cr26 | 5 | 5 | 5 | 5 |
| Cr27 | 5 | 5 | 5 | 5 |
| Cr28 | 5 | 5 | 5 | 5 |
| Cr29 | 5 | 5 | 5 | 5 |
| Cr30 | 5 | 5 | 5 | 5 |
| Cr31 | 5 | 5 | 5 | 5 |
| Cr32 | 5 | 5 | 5 | 5 |
| Cr33 | 5 | 5 | 5 | 5 |
| Cr34 | 5 | 5 | 5 | 5 |
| Cr35 | 5 | 5 | 5 | 5 |
| Cr36 | 5 | 5 | 5 | 5 |
| Cr37 | 5 | 5 | 5 | 5 |
| Overall performance | 5 | 5 | 5 | 4 |

- Cr20: Restaurant/Eating facilities (availability and quality)
- Cr21: Value for money of restaurant/eating facilities (prices)
- Cr22: Availability of bank/ATM facilities/money changers
- Cr23: Shopping facilities (availability and quality)
- Cr24: Value for money of shopping facilities (prices)

## 6.7 Prioritize criteria/aspects to be improved

**??** shows a prioritization regarding the improvement of criteria that appears in the rows according the categories (A, B, C, or D) in the columns."Blank" values in the the cells means that the mode of evaluations for the Group was equal to highest value possible in the scale. So, the recommendation is to "Mantain' the actions addressed to these criteria.

Analogous reasoning is applied to the criteria that does not appears in this Table. If a criteria is not listed in **??**, that is because its evaluation has a mode equal to the the highest evaluation in all categories.

The value "Priority" that appears in the cells means that the ATPs in the category should prioritize the improvement in this criteria to move the ATP to a´upper group. The values "Improve" means that there is a opportunity to improve the performance in that criteria, but it does not mean that it is enough to move the ATP to a upper groo.

## 7 Conclusion

This conclusion section is structured in two main paths of analysis: contribution to methodological improvements and for the particular case application.

Regarding the methodological aspects, the research described in this paper makes a contribution to fill a relevant methodological gap by proposing and applying a novel method for prioritizing aspects to be approached for increasing the level of satisfaction (LoS) that is purely non-compensatory and that takes a sensibility analysis of the credibility degree, as shown in Table 14. Complementing the comparison that appears in this table, we highlight the following properties of our proposal:

Table 14: Comparing our proposal against the state of art

|  | MAUT-based | Outranking | Our work |
|---|---|---|---|
| Deals with multiple criteria | Yes | Yes | Yes |
| Avoids compensatory effects intra criteria | No | Yes | Yes |
| Deal with multiple evaluators | Yes | Yes | Yes |
| Avoids compensatory effects introduced by multiple evaluator | No | No | Yes |
| Applied in a situation that considers evaluators using different set of criteria | No | No | Yes |
| Uses sensibility analysis of credibility degree to group instances | No | No | Yes |

1. It deals with situations where the product or service does not perform at the same level across all the factors or criteria that the sample looks at without adding any compensatory effects.
2. It enables many evaluators to assess a service using distinct criteria. This feature addresses the challenges associated with circumstances where the client, based on their own perception, does not feel adequately informed to assess a certain component of the service or product. The proposal deals with such situations without discarding evaluations or variables or even populating empty cells with a value, such as the average value of the responses provided by the other respondents.
3. By using a pure outranking algorithm, the system prevents compensatory effects from combining ratings from a group of people with different life experiences and expectations. For instance, if a customer rates a product or service as "Very good" and another user evaluates the same item as "Very poor," the evaluations are not combined as the service having a "Regular" or "Median" performance.

When applying the new proposal to evaluate the LoS in Brazilian ATPs, the modeling allowed us to conclude that:

In terms of the research's contribution to the specific subject of the application, we conclude that:

- This study fills a gap for a full non-compensatory method that is able to classify the LoS in the ATPs' context. Considering the perception of multiple evaluators and thousands of observations, the application demonstrated the feasibility of the proposal and covers aspects that previous modeling had not approached yet.

- The aspects to be prioritized for each group of ATPs are clearly shown in the paper so that the results can support decision-making by the ATP managers. Therefore, this study also contributes to supporting the planning of the ATP's infrastructure sector and to the continuous improvement of service levels provided to passengers by ATP's operators.
- It was shown that the proposal is a new way to evaluate Los in ATPs instead of using compensatory indices or measures on categorical scales. This is an important step forward in creating a multicriteria decision aid (MCDA) model to rate the level of services in ATPs.

As a limitation of the research, we mention the specific application of this research to Brazilian ATPs and the fact that the data collected does not comprise the importance of each criterion for each evaluator. These limitations only constrain the extension of the results from the specific application. In other words, the mentioned limitations are not a strong limitation under the methodological viewpoints and do not invalidate or imply a loss of generality of the method.

This is why new applications are suggested for future work while still following the structure of the proposal in different situations. This will help prove that it works and show if it could become a widely accepted method for judging LoS at airport passenger terminals.

Although the research was applied in the specific context of ATPs, it has the potential to be extended to other subjects, which should be explored in the next steps of the research.

# A  Previous modeling on LoS using MCDA methods

Table 15: Non-exausthive list of MAUT-based models for evaluating LoS

| Citation | Multi-criteria Method | Main Features |
|---|---|---|
| Wollmann et al. (2012) | AHP | Evaluates healthcare providers' service quality based on consumer perceptions. It applies AHP to analyse a cross-sectional survey conducted with 360 customers from seven health service providers in the Curitiba (Brazil) metropolitan area. |
| Drosos et al. (2015) | MUSA | Applies the MUSA method for evaluating quantitative global and partial satisfaction levels and eliciting the weak and strong points of service providers. |
| Jalali et al. (2016) | AHP | Uses AHP for constructing a n evaluation framework designed to order individuals in terms of their service provider loyalty. |
| Tlig and Rebai (2017) | TOPSIS | Combines the outputs from arithmetic operations based on fuzzy values with TOPSIS in a case study that evaluated and compared the LoS of five major airports located in North Africa. |
| Barbosa et al. (2018b) | AHP and TOPSIS | This study uses a global index, based on the compnation of AHP and TOPSIS, to assess the performance of power distribution utilities. The proposed method ranks service quality based on supply continuity, voltage compliance, and customer satisfaction. |
| Vahdat et al. (2019) | AHP and TOPSIS | Combines AHP, fuzzy set theory, and TOPSIS to rank hotel types based on the relative relevance of each SERVQUAL dimension in the business. |
| Drosos et al. (2020) | MUSA | Provides a multifaceted set of satisfaction, demanding, and improvement indices to support the analysis of the customer's satisfaction. |
| Vankova and Vavrek (2021) | TOPSIS | This research assesses residential social services provided to elderly people who cannot care for themselves. A sample composed by 519 senior residences in 77 Czech districts was evaluated by using TOPSIS integrated with the Coefficient of Variance approach, to objectively determine input indicator weights. |
| Panwar and Pant (2023) | AHP | This text outlines a model that combines Analytic Hierarchy Process (AHP) and Data Envelopment Analysis (DEA) to evaluate the performance of dormitories and mess facilities in a Higher Educational Institute (HEI). Analytic hierarchy process (AHP) is employed for qualitative analysis, whilst data envelopment analysis (DEA) is utilized for quantitative analysis. |
| Bozic et al. (2024) | MUSA | Analyzed passenger perceptions of air travel service quality, using data collected from 2016 up to 2018, regarding the services provided by Croatia main airport. |
| Golrizgashti et al. (2024) | DEMATEL | A combination of Delphi and DEMATEL is used to elicit causal linkages between criteria and sub-criteria to develop, assess, and prioritize cosmetic surgery clinic quality improvement criteria. |

Table 16: Non-exausthive list of non-compensatory models for evaluating LoS

| Citation | Multi-criteria Method | Main contributions |
|---|---|---|
| Freitas and Costa (1998) | ELECTRE III | Proposes the adoption of Electre III for sorting QoS. Such proposal, named by ELEQUAL, was applied to evaluate the services provided by car dealership, and was based in intorducing five artificial standard alternatives (A, B, C, D, E) together the service $X$ to be evaluated, each one having standardized performances in all criteria: A=Very good, B=Good, C=Neutral, D=Poor, and, E=Very poor. The classification of $X$ varies according to the position of $X$ in the ranking outputted from Electre III algorithm. |
| Costa et al. (2007) | ELECTRE TRI | Reviewed the proposal described in (Freitas and Costa 1998) by changing the Electre III algorithm by the pessimist or more exigent procedure of ELETRE TRI (Mousseau and Slowinski 1998). The main advantage of using Electre TRI was the reduction of the computational efforts, when compared with the use of Electre III, once there is no need to compare the standard alternatives (A,B,C,D,E) among them. As described in (Costa et al. 2007) the Electre TRI was applied to evaluate the customers' satisfaction regarding the services provided by a candy store |
| Lupo (2015) | ELECTRE III | Applied the Electre III for evaluating the quality of services provided by three international airports located in Sicilia, Italy. In this modeling AHP is used for eliciting the weights of the criteria, and, the average perceived quality on the $k^{th}$ service attribute was estimated by aggregating quality scores of related service attributes via the arithmetic mean operator mean. |
| Barbosa et al. (2018a) | Promethee II | Evaluates the performance of electricity distribution utilities with the use of a single global index by means of the Analytic Hierarchy Process (AHP) and Preference Ranking Organization Method for Enrichment Evaluations (PROMETHEE) methods. The proposed approach allows the ranking of service quality according to three dimensions: supply continuity, voltage conformity and customer satisfaction. While AHP was used to define the criteria weights, PROMETHEE presented the results in the form of a ranking, facilitating regulatory assessment of the distributors' performance, and thus improving the quality of services offered by utilities. |
| La Fata et al. (2019) | ELECTRE III | Applies a similar modeling to that described in Lupo (2015), in order to rank the quality of health services. It is observed that in this work AHP was used only in the phase of eliciting the weights and was not applied in the ranking phase. That is because, these works are classified here as having applied a non-compensatory ranking approach. |
| Sosyal et al. (2017) | Fuzzy PROMETHEE | Initially Fuzzy Analytic Hierarchy Process (FAHP) determines the significance of Parasuraman et al. (1988) e-service quality factors based on decision-making team surveys. Using customer survey data, the Fuzzy PROMETHEE approach evaluates airline websites' e-service quality. Four Turkish aviation businesses' websites were used in the investigation. |
| Tuzkaya et al. (2019) | PROMETHEE | In a real-life case study of an Istanbul public hospital, the proposal was used to evaluate service quality based on patient feedback. The evaluation provides thorough information on the impact of criteria and identifies possibilities to improve service quality through patient feedback analysis. |
| Yao et al. (2023) | Promethee II | Based on the results of a survey with 210 of experienced delivery platform users, four food deliveries platforms available at Twain were ranked by using a overall score from modified PROMETHEE by comparing actual performance to aspiration level. |

## B  The ELECTRE TRI ME algorithm based on Costa et al. (2018)

- Set $A \leftarrow \{a_1, a_2, \ldots a_m\}$ of $m$ instances to be evaluated.
- Set $E \leftarrow \{e_1, e_2, \ldots e_n\}$, composed by $n$ evaluators.
- Family $F$ of criteria. As each evaluator can have its own criteria set, them $F \leftarrow Fe_1 \cup Fe_2 \cup \ldots \cup Fe_n$, where:

    $Fe_1 \leftarrow \{k_{1e_1}, k_{2e_1}, \ldots k_{ve_1}\}$ is the subset composed by the $v$ criteria adopted by the evaluator $e_1$.
    $Fe_2 \leftarrow \{k_{1e_2}, k_{2e_2}, \ldots k_{xe_2}\}$ is the subset composed by the $x$ criteria adopted by the evaluator $e_2$.
    $Fe_j \leftarrow \{k_{1e_j}, k_{2e_j}, \ldots k_{e_2}\}$ is the subset composed by the $y$ criteria adopted by the evaluator $e_j$.
    $Fe_n \leftarrow \{k_{1e_n}, k_{2e_n}, \ldots k_{xe_n}\}$ is the subset composed by the $z$ criteria adopted by the evaluator $e_n$.
    $F \leftarrow Fe_1 \cup Fe_2 \cup \ldots \cup Fe_n$

- Weights (or constant of scales) of criteria $W \leftarrow We_1 \cup We_2 \cup \ldots \cup We_n$, where:

    $We_1 \leftarrow \{w_{1e_1}, w_{2e_1}, \ldots w_{ve_1}\}$ is the subset composed by the weights of $Fe_1$.
    $We_2 \leftarrow \{w_{1e_2}, w_{2e_2}, \ldots w_{xe_2}\}$ is the subset composed by the weights of $Fe_2$.
    $We_j \leftarrow \{w_{1e_j}, w_{2e_j}, \ldots w_{e_2}\}$ is the subset composed by the weights of $Fe_j$.
    $We_n \leftarrow \{w_{1e_n}, w_{2e_n}, \ldots w_{xe_n}\}$ is the subset composed by the weights of $Fe_n$.

- Performance of alternatives
  $G \leftarrow Ge_1(a_1) \cup \ldots Ge_j(a_i) \cup \ldots \cup Ge_n(a_m)$, where:

    $Ge_1(a_1) \leftarrow \{g_{1e_1}(a_1), g_{2e_1}(a_1), \ldots g_{ve_1}(a_1)\}$ is the subset composed by the performance of alternative $a_1$ under the perspective of the evaluator $e_1$ and the set of criteria $Fe_1$. For example, $g_{2e1}(a_i)$ means the performance of alternative $a_i$ under criterion 2 and under the viewpoint of the evaluator $e_1$.
    $ge_j(a_i) \leftarrow \{g_{1e_j}(a_i), g_{2e_j}(a_i), \ldots g_{|Fe_j|e_j}(a_i)\}$ is the subset composed by the performance of alternative $a_i$ under the perspective of the evaluator $e_j$ and the set of criteria $Fe_j$.
    $\ldots$
    $ge_n(a_m) \leftarrow \{g_{1e_n}(a_m), \ldots, g_{|Fe_j|e_n}(a_m)\}$ is the subset composed by the performance of alternative $a_m$ under the perspective of the evaluator $e_n$ and the set of criteria $Fe_n$.

- Set of categories $C \leftarrow \{C_1, C_2, \ldots C_k\}$ in which the instances will be sorted.

    $C$ is ranked from the best to the worst, i.e. $C_1$ is the best category and $C_k$ is the worst one. The number of categories is the same for all $e \in E$ and all $Fe \in F$.

    Set of categories' boundaries $B$. Notice that $|B|$ is equal to $|C| - 1$, and that $b_{pfe}$ is the boundary of the $p^{th}$ category, in the $f^{th}$ criterion for the $e^{th}$ evaluator. For example, $b_{123}$ is the boundary of category 1, in criterion 2 for evaluator 3.

    $B \leftarrow Be_1 \cup Be_2 \cup \ldots \cup Be_n$, where:

- Cut-level $\lambda$ that means the minimal credibility degree accepted for the sorting.

[*Step 1:* ] Input $A, E, F, W, G, C, B$, and, $\lambda$

[*Step 2:* ] Compute the overall credibility degree: $\sigma(a_i, b_p)$. In other words, calculate the level of credibility with the assertive: "$a_i$ is at least as good as the boundary $b_p$.

for $i = 1$ to $|A|$ do
  for $p = 1$ to $|C| - 1$ do
    for $j = 1$ to $|F|$ do
    $\sigma(a_i, b_p) \leftarrow \left[\dfrac{1}{\sum_{j=1}^{|F|}}\right] * \sum_{j=1}^{|F|} w_j * c_j(a_i, b_{p,j})$
    Where:
    $c_j(a_i, b_{p,j}) \leftarrow \begin{cases} 1, & \iff g_j(a_i) \geq b_{p,j} \\ 0, & otherwise \end{cases}$

[*Step 3:* ] Run the more exigent sorting procedure of ELECTRE TRI Mousseau and Slowinski (1998) to obtain objects clustered.

The choice to use the more exigent procedure is because it produces a sorting more suitable to the customer enrollment, while the more benevolent procedure is more close to the service provider enrollment.

```
for i = 1 to |A| do
    p ← 1
    for p = 1 to |C| − 1 do
        if σ(a_i, b_p) ≥ λ
            then
                assign a_i to category C_p
                p ← |C| + 1
            else
                p ← p + 1
                if p = |C|
                    then
                        assign a_i to category C_p
```
$\lambda$ means the minimum value of credibility required.